\documentclass[conference]{IEEEtran}
\IEEEoverridecommandlockouts
\usepackage{cite}
\usepackage{amsmath,amssymb,amsfonts}
\usepackage{graphicx}
\usepackage{textcomp}
\usepackage{xcolor}

\usepackage{graphicx}
\usepackage{dcolumn}
\usepackage{bm}
\usepackage{xcolor}
\usepackage{qcircuit}
\usepackage{braket}
\usepackage{float}
\usepackage{algorithm}
\usepackage{algorithmicx}
\usepackage{algpseudocode}
\usepackage{mathtools}
\usepackage{hyperref}
\usepackage{makecell}

\def\BibTeX{{\rm B\kern-.05em{\sc i\kern-.025em b}\kern-.08em
    T\kern-.1667em\lower.7ex\hbox{E}\kern-.125emX}}

\renewcommand{\figureautorefname}{Figure~\negthinspace}

\renewcommand{\tableautorefname}{Table~\negthinspace}
\renewcommand{\sectionautorefname}{Section~\negthinspace}

\begin{document}

\title{Differentiable Quantum Architecture Search in Asynchronous Quantum Reinforcement Learning
\thanks{The views expressed in this article are those of the authors and do not represent the views of Wells Fargo. This article is for informational purposes only. Nothing contained in this article should be construed as investment advice. Wells Fargo makes no express or implied warranties and expressly disclaims all legal, tax, and accounting implications related to this article.}
}

\author{\IEEEauthorblockN{ Samuel Yen-Chi Chen}
\IEEEauthorblockA{
\textit{Wells Fargo}\\
New York, NY, USA \\
yen-chi.chen@wellsfargo.com}
}

\maketitle

\begin{abstract}
The emergence of quantum reinforcement learning (QRL) is propelled by advancements in quantum computing (QC) and machine learning (ML), particularly through quantum neural networks (QNN) built on variational quantum circuits (VQC). These advancements have proven successful in addressing sequential decision-making tasks. However, constructing effective QRL models demands significant expertise due to challenges in designing quantum circuit architectures, including data encoding and parameterized circuits, which profoundly influence model performance. In this paper, we propose addressing this challenge with differentiable quantum architecture search (DiffQAS), enabling trainable circuit parameters and structure weights using gradient-based optimization. Furthermore, we enhance training efficiency through asynchronous reinforcement learning (RL) methods facilitating parallel training. Through numerical simulations, we demonstrate that our proposed DiffQAS-QRL approach achieves performance comparable to manually-crafted circuit architectures across considered environments, showcasing stability across diverse scenarios. This methodology offers a pathway for designing QRL models without extensive quantum knowledge, ensuring robust performance and fostering broader application of QRL.
\end{abstract}


\section{\label{sec:Indroduction}Introduction}

Quantum computing (QC) theoretically possesses the potential to fundamentally transform computational tasks, presenting clear advantages over classical computers \cite{nielsen2010quantum}.
The advancement in QC hardware, coupled with classical ML techniques, enables the progression of quantum machine learning (QML). This is realized through the hybrid quantum-classical computing paradigm \cite{cerezo2021variational, bharti2022noisy}, wherein both classical and quantum computers are harnessed. Specifically, computational tasks suited for QC capabilities are executed on quantum computers, while tasks such as gradient calculations, well-handled by classical computers, remain within their domain.
The variational quantum circuit (VQC) serves as the fundamental component in existing QML methodologies. A plethora of QML models, based on VQC, have been developed to address various machine learning (ML) tasks including classification \cite{mitarai2018quantum,chen2021end,chen2022quantumCNN,oh2020tutorial,qi2023qtnvqc,wu2022poster}, time-series prediction \cite{chen2022quantumLSTM,chen2022reservoir,bausch2020recurrent}, generative modeling \cite{chu2023iqgan,stein2021qugan}, natural language processing \cite{yang2021decentralizing,li2023pqlm,yang2022bert,di2022dawn,stein2023applying}, and reinforcement learning \cite{chen2023quantum_LSTM_RL,chen2023efficientQRL_QRC,chen2022variationalQRL,chen19,lockwood2020reinforcement,skolik2021quantum,jerbi2021variational}.
Despite the achievements of various QML models, a significant challenge hindering widespread adoption is the necessity for extensive expertise in designing effective quantum circuit architectures. For instance, crafting the structure of encoding and parameterized circuits within the VQC demands specialized design considerations, including appropriate entanglement to showcase quantum advantages. There exists a pressing demand for an automated procedure capable of streamlining the search for high-performing quantum circuit architectures.
In this manuscript, we address this challenge through the implementation of differentiable quantum architecture search (DiffQAS). Our objective is to integrate DiffQAS into quantum reinforcement learning (QRL), as reinforcement learning (RL) stands out in the realm of ML for its handling of sequential decision-making problems and potential to exhibit high-level problem-solving capabilities. Specifically, we examine a selection of VQC block candidates and assign trainable \emph{structural weights} to these blocks. Through gradient descent optimization, we concurrently learn these structural weights alongside the conventional quantum circuit parameters (rotation angles). Furthermore, departing from prior works in QAS for QRL, we train the agents using asynchronous training rather than single-process policy updates, thereby leveraging the computational resources of multi-core CPUs or potentially multiple quantum processing units (QPUs) in the future.
Numerical simulations are employed to illustrate the efficacy of the proposed DiffQAS-QRL framework in identifying VQC architectures capable of achieving high scores in diverse testing environments. Specifically, we demonstrate the stability of the proposed method across various environments, contrasting with the inconsistent performance of manually-designed architectures across different scenarios. This observation underscores the necessity for a task-agnostic automated procedure in designing QRL circuits and underscores the effectiveness of the proposed DiffQAS-QRL framework.

This manuscript is structured as follows: \sectionautorefname{\ref{sec:Related_Work}} presents a concise overview of the current advancements in QAS and QRL. In \sectionautorefname{\ref{sec:QRL}}, the fundamental concepts of quantum RL and VQC are elucidated, forming the foundational components of extant QML and QRL models, which represent the focus of inquiry for the proposed framework. The formulation of the differentiable QAS problem is delineated in \sectionautorefname{\ref{sec:DiffQAS}}, while the intricacies of the proposed DiffQAS-QRL framework are expounded upon in \sectionautorefname{\ref{sec:Methods}}, along with the methodologies for numerical simulation and the corresponding outcomes discussed in \sectionautorefname{\ref{sec:Experiments}}. Ultimately, the findings are summarized in \sectionautorefname{\ref{sec:Conclusion}}.

\begin{figure}[htbp]
\begin{center}
\includegraphics[width=1\columnwidth]{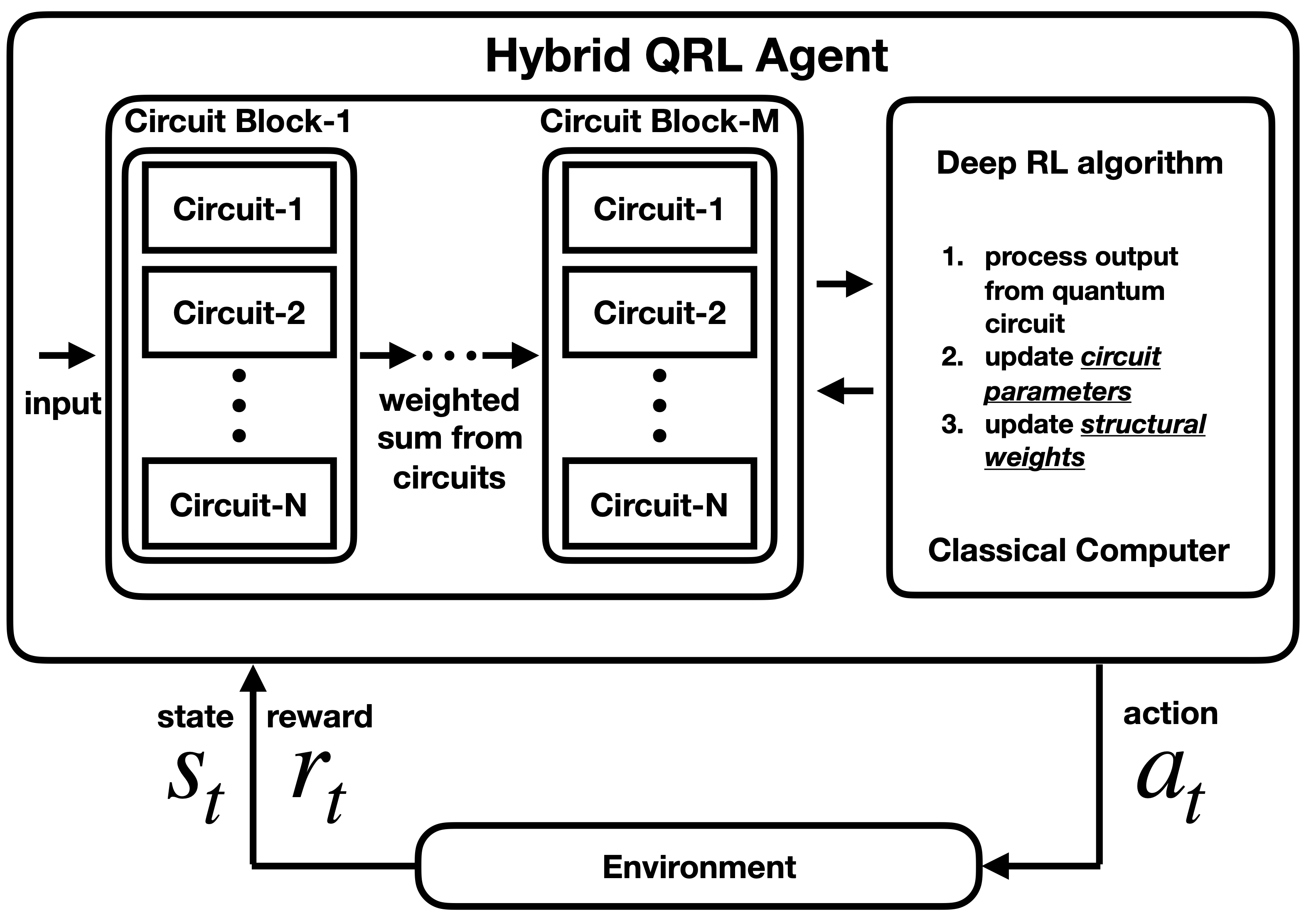}
\caption{{\bfseries Differentiable QAS for QRL framework.}}
\label{fig:DiffQAS_QRL_framework}
\end{center}
\end{figure}

\section{\label{sec:Related_Work}Related Work}
Quantum reinforcement learning (QRL) has been a subject of exploration since the groundbreaking work by Dong et al. in 2008 \cite{dong2008quantum}. Initially, its practicality was hindered by the requirement to construct environments entirely in a quantum fashion, thus limiting its real-world applicability. However, subsequent advancements in QRL, leveraging variational quantum circuits (VQCs), have broadened its horizons to encompass classical environments with both discrete \cite{chen19} and continuous observation spaces \cite{lockwood2020reinforcement, skolik2021quantum}. The evolution of QRL has witnessed performance improvements through the adoption of policy-based learning methodologies, including Proximal Policy Optimization (PPO) \cite{hsiao2022unentangled}, Soft Actor-Critic (SAC) \cite{lan2021variational}, REINFORCE \cite{jerbi2021variational}, Advantage Actor-Critic (A2C) \cite{kolle2024quantum}, and Asynchronous Advantage Actor-Critic (A3C) \cite{CHEN2023321Async}. Moreover, in addressing the challenges presented by partially observable environments, researchers have explored the utilization of quantum recurrent neural networks such as quantum LSTM as RL policies \cite{chen2023quantum_LSTM_RL,chen2023efficientQRL_QRC}. 
Recent advancements also encompass hybrid models, wherein a classical neural network is trained to dynamically adjust the parameters of the quantum circuit. This enables the model to address intricate sequential decision-making tasks without relying on quantum recurrence \cite{chen2024learning}.

Nevertheless, the accomplishments mentioned in QRL necessitate profound expertise in designing high-performing quantum circuit architectures to leverage potential quantum advantages. Consequently, there is an imminent requirement to develop automated procedures for designing quantum circuit architectures to cater to the demands of various application domains.
Machine learning techniques have been employed to address various challenges in quantum computing, such as quantum architecture search (QAS). The objectives of QAS may include generating desired quantum states \cite{kuo2021quantum,ye2021quantum,kimura2022quantum,sogabe2022model,lu2023qas,kundu2024enhancing,sunkel2023ga4qco,zhu2023quantum,chen2023QRL_QAS,selig2023deepqprep,sun2024quantum}, discovering efficient circuits for solving chemical ground states \cite{ostaszewski2021reinforcement,wang2023automated,he2023gnn,sun2024quantum,deng2023progressive}, addressing optimization tasks \cite{yao2022monte,duong2022quantum,wang2023automated,wu2023quantumdarts,sun2024quantum,zhang2022differentiable,sun2024differentiable}, optimizing quantum circuits for specific hardware architectures \cite{fosel2021quantum}, compiling circuits \cite{he2022quantum,he2022search,chen2022efficient}, or conducting machine learning tasks \cite{ding2022evolutionary,duong2022quantum,wu2023quantumdarts,zhang2023evolutionary,ding2023multi,subasi2023toward,sun2023differentiable,zhang2021neural,du2022quantum}.
Various methodologies are employed to discover the optimal circuit for specific tasks. For instance, reinforcement learning-based approaches are explored in works such as \cite{kuo2021quantum,ye2021quantum,fosel2021quantum,ostaszewski2021reinforcement,yao2022monte,kimura2022quantum,sogabe2022model,kundu2024enhancing,zhu2023quantum,chen2023QRL_QAS,chen2022efficient}, while different variants of evolutionary algorithms are utilized in works like \cite{ding2022evolutionary,zhang2023evolutionary,ding2023multi,sunkel2023ga4qco} to search for circuits. Additionally, differentiable QAS methods have been developed to leverage gradient-based techniques effectively \cite{wu2023quantumdarts,zhang2022differentiable,sun2024differentiable,sun2023differentiable}. Various approaches to encode quantum circuit architecture have been proposed, with some utilizing graph-based methods as seen in works such as \cite{duong2022quantum,he2023gnn}, while others, like \cite{fosel2021quantum}, consider convolutional neural network-based methods. As for circuit performance metrics, they may involve direct evaluation of circuit performance on specific tasks \cite{ostaszewski2021reinforcement,ding2022evolutionary,wang2023automated}, or assessing the proximity of the generated circuit to the actual circuit \cite{kuo2021quantum,ye2021quantum,duong2022quantum}. To alleviate the computational resources needed for direct evaluation, predictor-based methods have been proposed, employing neural networks to predict quantum model performance without direct circuit evaluation \cite{deng2023progressive,zhang2021neural}.

The proposed framework in this paper extends the asynchronous training of QRL described in the work \cite{CHEN2023321Async,chen2023efficientQRL_QRC,chen2024learning} to include the capability of differentiable QAS to search for the best performing circuit. Our work also differentiates from the previous work \cite{sun2023differentiable} as our work considers the asynchronous training which can leverage parallel computing resource or in the future, multiple-QPU environments.
Our work is also different from the previous work such as \cite{ding2022evolutionary} since our work leverages the differentiable method, unlike the evolutionary methods requiring a large amount of performance evaluation. 
\section{\label{sec:QRL}Quantum Reinforcement Learning}
\subsection{\label{sec:VQC}Variational Quantum Circuits}
\begin{figure}[htbp]
\begin{center}
\includegraphics[width=1\columnwidth]{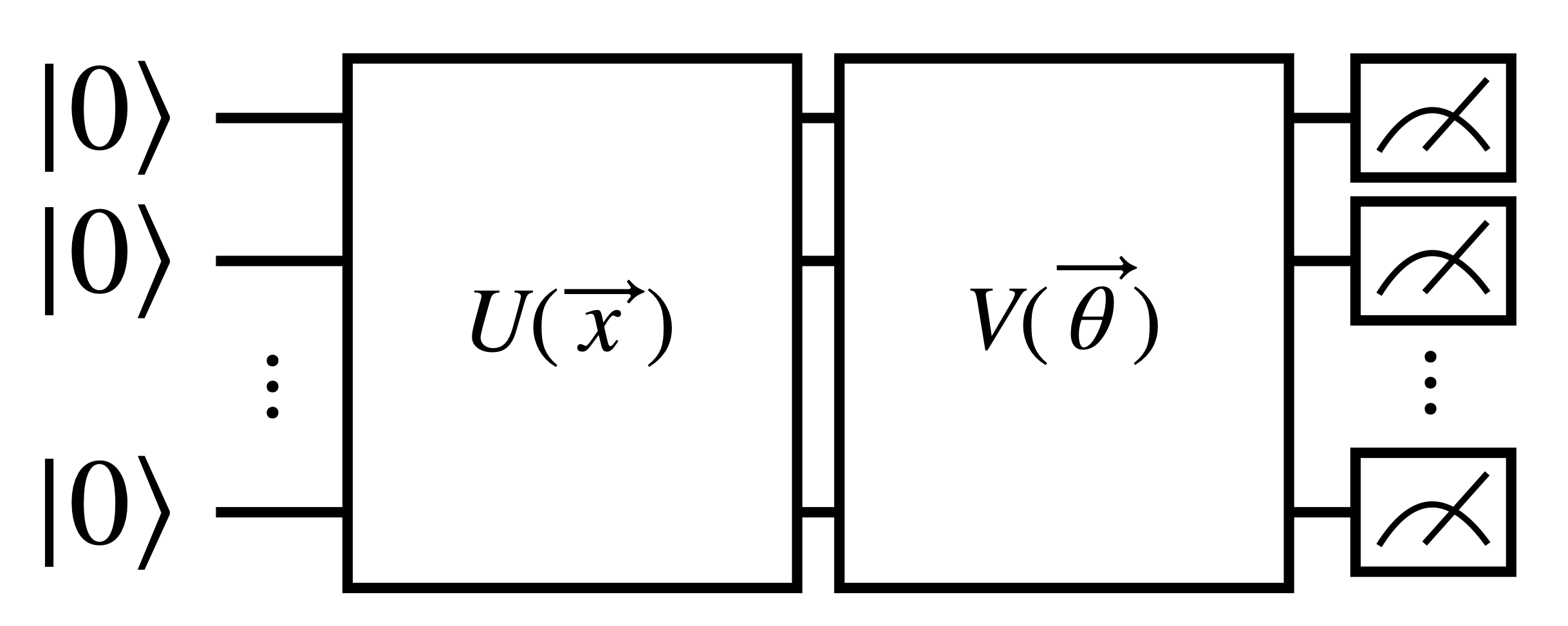}
\caption{{\bfseries Generic structure of a VQC.}}
\label{fig:generic_vqc}
\end{center}
\end{figure}

Quantum machine learning largely depends on the trainable quantum circuits: variational quantum circuit (VQC), also known as parameterized quantum circuits (PQC). The VQC, as shown in \figureautorefname{\ref{fig:generic_vqc}}, usually has three basic components: \emph{encoding} circuit $U(\vec{x})$, \emph{variational} circuit $V(\vec{\theta})$ and the final \emph{measurement} part. 
The purpose of encoding circuit $U(\vec{x})$ is to transform the input vector $\vec{x}$ into a quantum state $U(\vec{x})\ket{0}^{\otimes n}$, where $\ket{0}^{\otimes n}$ is the ground state of the quantum system and $n$ represents the number of the qubit. The encoded state then go through the variational circuit and becomes $V(\vec{\theta})U(\vec{x})\ket{0}^{\otimes n}$. To retrieve the information from the VQC, measurements can be carried out with pre-defined observables $\hat{B}_{k}$. The VQC operation can be seen as as quantum function $\overrightarrow{f(\vec{x} ; \vec{\theta})}=\left(\left\langle\hat{B}_1\right\rangle, \cdots,\left\langle\hat{B}_n\right\rangle\right)$, where $\left\langle\hat{B}_{k}\right\rangle =\left\langle 0\left|U^{\dagger}(\vec{x})V^{\dagger}(\vec{\theta}) \hat{B}_{k} V(\vec{\theta})U(\vec{x})\right| 0\right\rangle$.
Expectation values $\left\langle\hat{B}_{k}\right\rangle$ can be obtained by performing multiple samplings (shots) on actual quantum hardware or through direct computation when utilizing simulation software.

\subsection{Quantum RL}
\emph{Reinforcement learning} (RL) constitutes a ML paradigm wherein an \emph{agent} endeavors to achieve a predefined objective or goal through interactions with an \emph{environment} $\mathcal{E}$ within discrete time intervals~\cite{sutton2018reinforcement}. At each time step $t$, the agent perceives a \emph{state} $s_t$ and subsequently selects an \emph{action} $a_t$ from the action space $\mathcal{A}$ based on its prevailing \emph{policy} $\pi$. The policy signifies a mapping from a specific state $s_t$ to the probabilities associated with selecting an action from $\mathcal{A}$. Upon executing action $a_t$, the agent receives a scalar \emph{reward} $r_t$ and the updated subsequent state $s_{t+1}$ of the environment. For episodic tasks, this sequence of actions and new states recurs over multiple time steps until the agent either reaches a terminal state or exhausts the allowable number of steps.

A category of RL training algorithms referred to as \emph{policy gradient} methods centers on optimizing the policy function, represented as $\pi(a|s;\theta)$, which is parameterized by $\theta$. These parameters, denoted as $\theta$, undergo updates via a gradient ascent process on the expected total return, $\mathbb{E}[R_{t}]$.
In classical RL, the function $\pi(a|s;\theta)$ is realized using a deep neural network (DNN), with $\theta$ representing the DNN's weights and biases. In quantum RL, the policy function $\pi(a|s;\theta)$ can be implemented through VQCs or hybrid models that integrate both VQCs and conventional DNNs. The integration of VQCs and DNNs forms a directed acyclic graph (DAG) and the whole model can be optimized in an end-to-end manner \cite{chen2022variationalQRL,chen2023quantum_LSTM_RL,lockwood2020reinforcement,chen19,skolik2021quantum,jerbi2021variational}.
An exemplary instance of a policy gradient algorithm is the REINFORCE algorithm, initially proposed in~\cite{williams1992simple}. Within the conventional REINFORCE algorithm, the parameters $\theta$ undergo updates in the direction of $\nabla_{\theta} \log \pi\left(a_{t} | s_{t} ; \theta\right) R_{t}$, constituting an unbiased estimate of $\nabla_{\theta} \mathbb{E}\left[R_{t}\right]$.

Nonetheless, the policy gradient estimate frequently encounters high variance, presenting challenges during training. To mitigate this variance while preserving its unbiased nature, practitioners often subtract a term referred to as the \emph{baseline} from the return. This baseline, denoted as $b_{t}(s_{t})$, constitutes a learned function of the state $s_{t}$. Consequently, the updated expression becomes $\nabla_{\theta} \log \pi\left(a_{t} | s_{t} ; \theta\right)\left(R_{t}-b_{t}\left(s_{t}\right)\right)$.
In policy gradient RL, a prevalent selection for the baseline $b_t(s_t)$ is an estimation of the value function $V^\pi(s_t)$. Employing this baseline choice often yields a policy gradient estimate with reduced variance \cite{sutton2018reinforcement}. The difference $R_t - b_t = Q(s_t, a_t) - V(s_t)$ can be interpreted as the \emph{advantage} $A(s_t, a_t)$ of action $a_t$ at state $s_t$. Conceptually, the advantage represents the relative "goodness or badness" of action $a_t$ concerning the average value at state $s_t$. This approach is known as the advantage actor-critic (A2C) method.
The quantum version of REINFORCE algorithm with value function baselines is described in the work \cite{jerbi2021variational}. The work \cite{jerbi2021variational} further demonstrates the advantage of hybrid quantum-classical RL over classical models on discrete logarithm problem.

The asynchronous advantage actor-critic (A3C) algorithm \cite{mnih2016asynchronous} extends the A2C approach by employing multiple concurrent actors to learn the policy through parallelization. Asynchronous training of RL agents entails running multiple agents concurrently on various instances of the environment, enabling them to encounter diverse states at each time step. This reduced correlation between states or observations enhances the numerical stability of on-policy RL algorithms like actor-critic \cite{mnih2016asynchronous}. Moreover, asynchronous training obviates the need for maintaining an extensive replay memory, thereby reducing memory requirements \cite{mnih2016asynchronous}. Recent investigations \cite{CHEN2023321Async} have indicated that the quantum version of A3C can yield superior performance in specific benchmark tasks compared to their classical counterparts under certain conditions. Asynchronous QRL can facilitate efficient training, leveraging multiple CPU cores or QPU cores, contingent on whether a simulation backend or a real quantum device is employed.
\begin{figure}[htbp]
\begin{center}
\includegraphics[width=1\columnwidth]{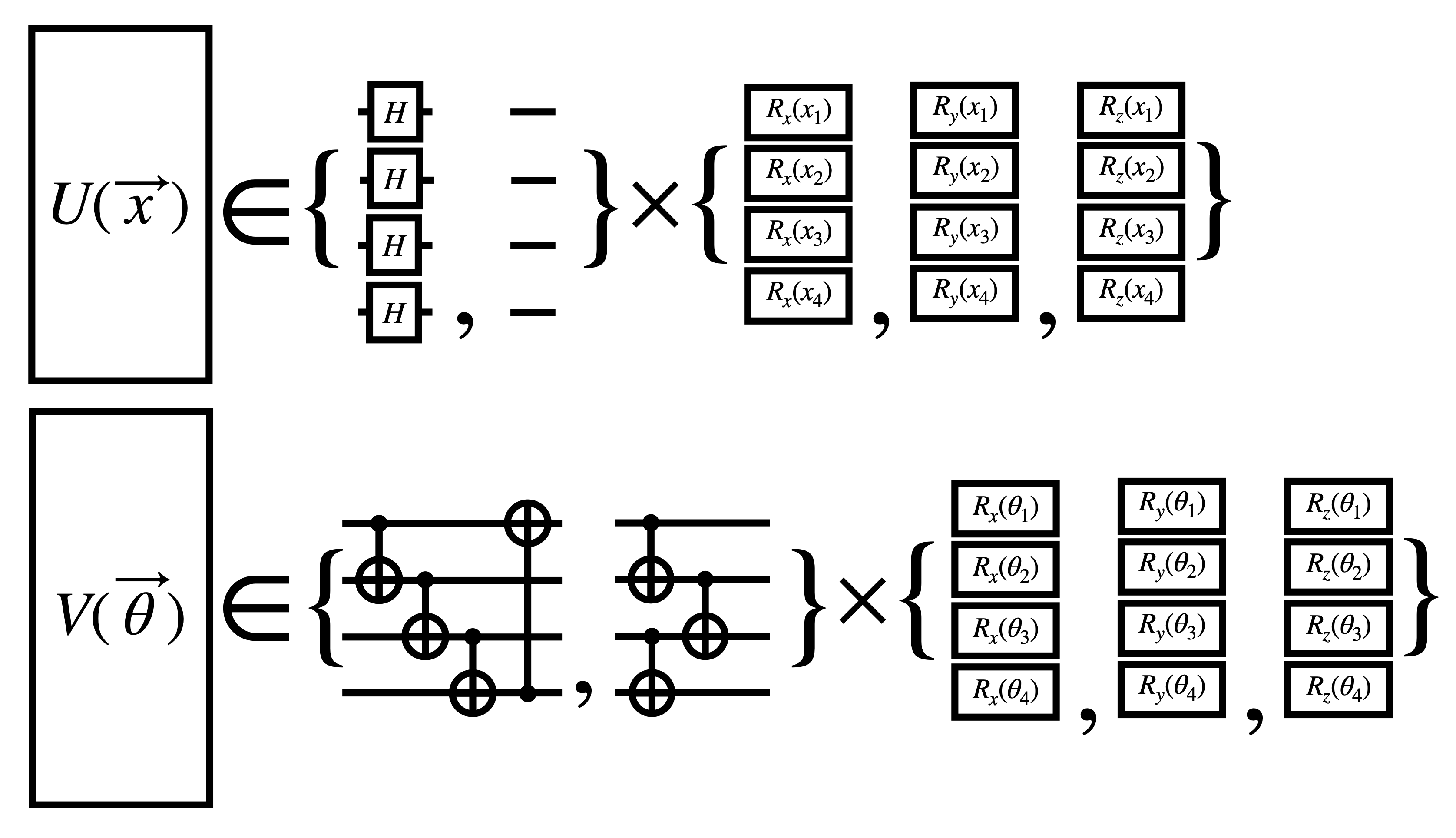}
\caption{{\bfseries Ansatzes of VQC considered in this work.}}
\label{fig:vqc_ansatz_candidate}
\end{center}
\end{figure}
\section{\label{sec:DiffQAS}Differentiable Quantum Architecture Search}
Drawing inspiration from classical neural architecture search (NAS) \cite{liu2018darts} and foundational research on differentiable QAS \cite{zhang2022differentiable}, our DiffQAS approach commences with a collection of candidate subcircuits. Suppose we aim to construct a quantum circuit $\mathcal{C}$ requiring several sub-components $\mathcal{S}_{1}, \mathcal{S}_{2}, \cdots, \mathcal{S}_{n}$. Each $\mathcal{S}_{i}$ is associated with a corresponding set of allowable circuit choices $\mathcal{B}_{i}$, where $|\mathcal{B}_{i}|$ denotes the number of permissible circuit choices for each sub-component $i$. Thus, the total number of potential outcomes for circuit $\mathcal{C}$ is given by $N = |\mathcal{B}_{1}| \times |\mathcal{B}_{2}| \times \cdots \times |\mathcal{B}_{n}|$. Structural weights $w_{j}$, where $j \in \{1, \cdots, N\}$, are assigned to each possible circuit realization $\mathcal{C}_{j}$. Additionally, we assume that each $\mathcal{C}_{j}$ possesses its own trainable parameters $\theta_{j}$.

Consider a ML task for which certain circuit realizations $\mathcal{C}_{j}$, when trained with their corresponding parameters $\theta_{j}$, may offer viable approaches. However, each specific circuit $\mathcal{C}_{j}$ does not guarantee an optimal solution; it may yield optimal, suboptimal, or even ineffective outcomes. We introduce the ensemble function $f_{\mathcal{C}}$, defined as the weighted sum of all potential circuit realizations, denoted as $f_{\mathcal{C}} = \sum_{j = 1}^{N} w_{j}f_{\mathcal{C}_{j}}$. For clarity, we omit the notation for quantum circuit parameters (rotation angles) $\theta_{j}$ and input vector $\vec{x}$. Subsequently, the output from the ensemble function $f_{\mathcal{C}}$ is subjected to processing by the loss function $\mathcal{L}(f_{\mathcal{C}})$. Utilizing automatic differentiation algorithms, the gradient with respect to the structural weights $w_{j}$ can be computed as $\nabla_{w_{j}} \mathcal{L}(f_{\mathcal{C}})$. Conventional gradient-based optimizers can then be employed to optimize the weights $w_{j}$.

\begin{figure}[htbp]
\begin{center}
\includegraphics[width=1\columnwidth]{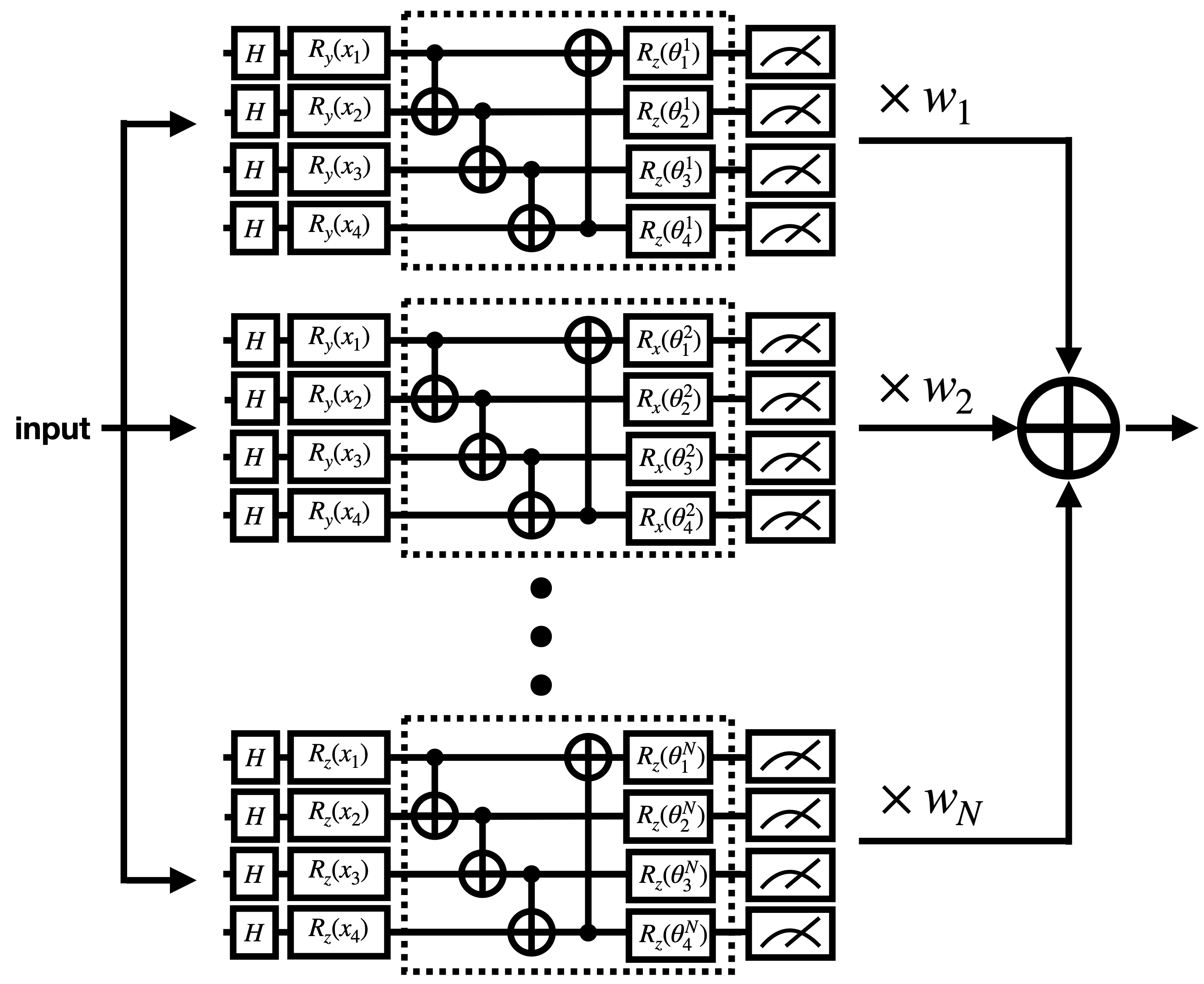}
\caption{{\bfseries Quantum function with differentiable structure weights.}}
\label{fig:DiffStructure_VQC}
\end{center}
\end{figure}

In the proposed DiffQAS framework, VQCs can be assembled from a set of ansatzes, as illustrated in \figureautorefname{\ref{fig:vqc_ansatz_candidate}}. Specifically, for the encoding circuit $U(\vec{x})$, it may incorporate or exclude the Hadamard gate, followed by the application of one of the rotation gates ($R_{x}$, $R_{y}$, and $R_{z}$). Regarding the variational circuit $V(\vec{\theta})$, two options exist for the entanglement and three for the parameterized rotation, resulting in six candidates for both the encoding and variational components. Consequently, for a single VQC, there are $6 \times 6 = 36$ potential configurations. Enumerating all $N = 36$ feasible circuit realizations, we assign structural weights $w_{j}$ to each, as depicted in \figureautorefname{\ref{fig:DiffStructure_VQC}}. To meet the requirements of the given task, we can stack multiple ensemble functions $f_{\mathcal{C}}$ to construct deep quantum circuits, as demonstrated in \figureautorefname{\ref{fig:QA3C_QRL}}. It should be noted that constructing deep quantum circuits increases the number of structural weights $w_{j}$. For instance, if we build deep quantum circuits with $M$ circuit blocks, there will be $N \times M$ structural weights, where $N$ represents the number of possible circuit realizations. The structural weights across all layers can be optimized using gradient-based optimizers. 
\section{\label{sec:Methods}Methods}
\subsection{\label{sec:DiffQAS_QRL} Differentiable QAS for QRL}
\begin{figure}[htbp]
\begin{center}
\includegraphics[width=1\columnwidth]{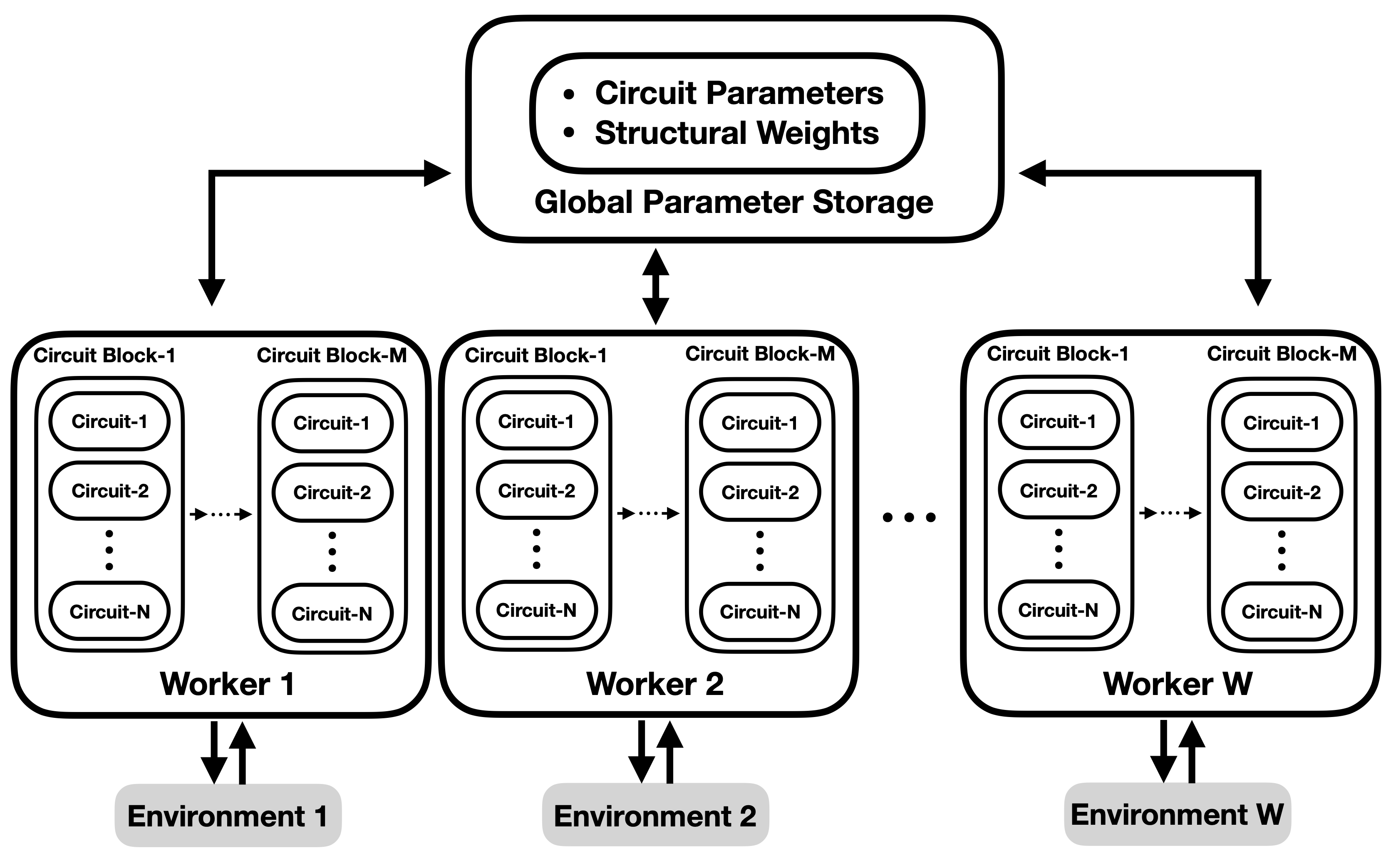}
\caption{{\bfseries Asynchronous quantum RL with differentiable QAS.}}
\label{fig:QA3C_QRL}
\end{center}
\end{figure}

In QRL, certain functions such as \emph{value function} and \emph{policy function} would be implemented by hybrid quantum-classical models. Inside the hybrid models, the quantum components are realized via VQCs as described in \sectionautorefname{\ref{sec:VQC}}. Traditionally, the architecture of VQC is designed before the model training process. While this method has shown several successful QRL applications \cite{chen19,lockwood2020reinforcement,skolik2021quantum,jerbi2021variational}, it is with certain limitations. For example, the designed architecture may be suitable for only a small set of tasks. And the design of new architecture may require domain expertise. In this paper, we relax some of the constraints via defining the hybrid models with the \emph{trainable structural weights}.
Consider the value function $Q(\vec{s};\Theta)$, where $\vec{s}$ is the state or observation from the environment and $\Theta$ is the whole parameter set including classical and quantum ones. The value function can then be expressed as $Q(\vec{s};\Theta) = G_{\eta} \circ F_{\theta} \circ H_{\delta}(\vec{s}) $ where $\eta, \theta, \delta \in \Theta$ are trainable parameters and $G$ and $H$ are classical functions and $F$ is the quantum function. 
The quantum function $F(\vec{x};\theta)$ may be composed of an array of candidate functions $f_{i}(\vec{x};\theta_{i})$ with trainable structural weights $w_{i}$. It can be expressed as $F(\vec{x};\theta) = \sum w_{i} f_{i}(\vec{x};\theta_{i})$.
The asynchronous training of the quantum model is presented in \figureautorefname{\ref{fig:QA3C_QRL}}. 
In conventional quantum A3C \cite{CHEN2023321Async}, the architectures of VQC models remain fixed, with only the gradients of quantum circuit parameters transmitted to the central storage. In the proposed DiffQAS with A3C, however, both the gradients of quantum circuit parameters and the gradients of structural weights are uploaded.
\section{\label{sec:Experiments}Experiments}
In this study, we utilize the following open-source tools for simulation purposes. We employ PennyLane~\cite{bergholm2018pennylane} for quantum circuit construction and PyTorch for developing the overarching hybrid quantum-classical model.
The hyperparameters for the proposed DiffQAS in RL with QA3C training \cite{CHEN2023321Async,chen2023efficientQRL_QRC} are set as follows: Adam optimizer with a learning rate of $1 \times 10^{-4}$, $beta_{1} = 0.92$, $beta_{2} = 0.999$, $L = 5$ for model lookup steps, and a discount factor $\gamma = 0.9$. During the asynchronous training process, local agents or models compute their gradients every $L$ steps, corresponding to the trajectory length used during model updates. The number of parallel processes (number of local agents) is $80$. To illustrate both the trend and stability, we present results with the average score alongside its standard deviation over the past 5,000 episodes. The standard deviation is shown as the shaded area in the result plots. We summarize the VQC baselines in the \tableautorefname{\ref{tab:different_circuit_baselines}}. Each VQC configuration consists of an $8$-qubit system and $2$ variational layers, resulting in a total of $16$ trainable quantum parameters ($2 \times 8 = 16$).
\subsection{Environment}
\begin{figure}[htbp]
\begin{center}
\includegraphics[width=1\columnwidth]{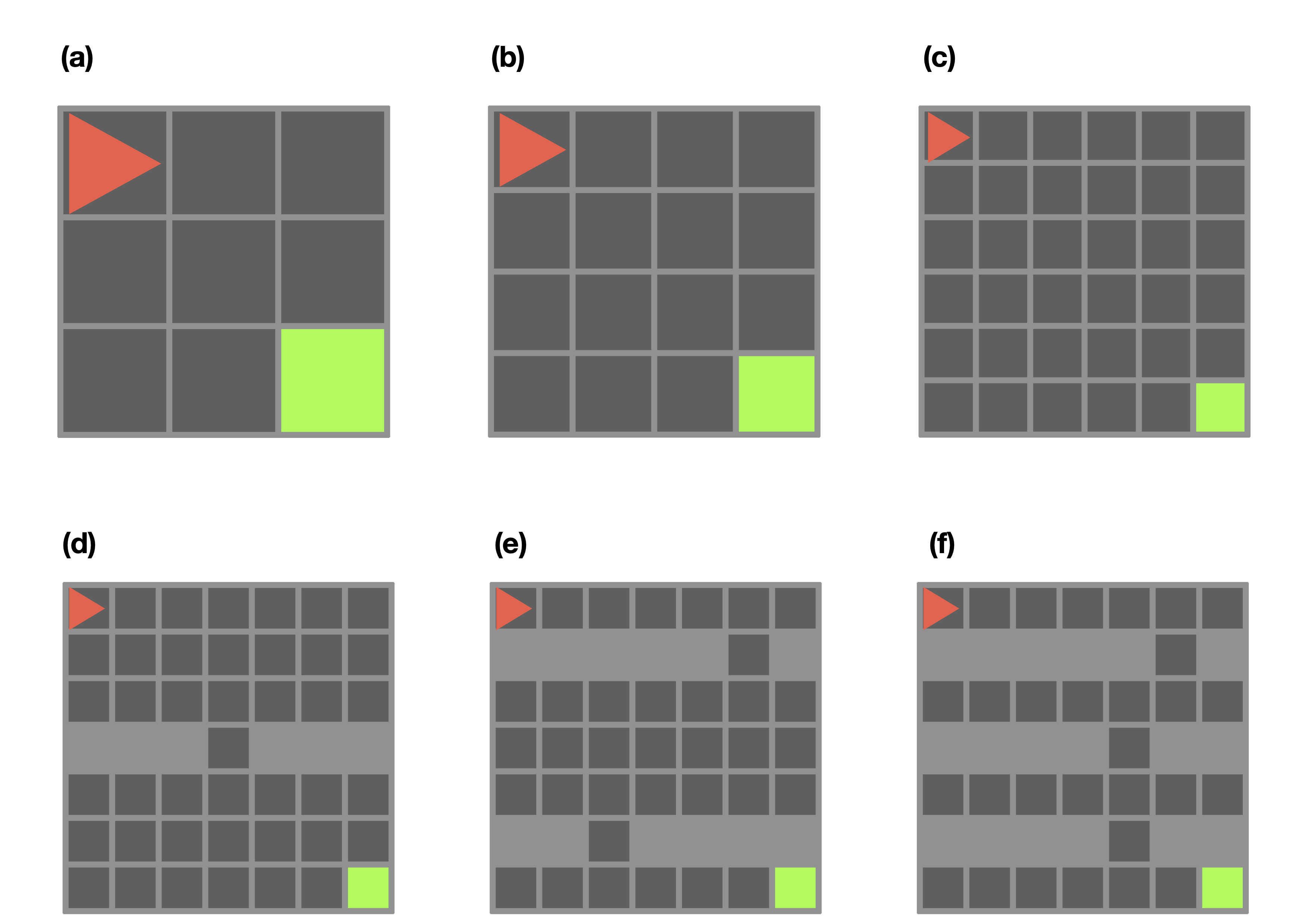}
\caption{{\bfseries MiniGrid Environments.}}
\label{fig:minigrid_envs}
\end{center}
\end{figure}

The MiniGrid environment \cite{gym_minigrid} presents a more complex scenario, featuring a significantly larger observation input for the quantum RL agent. In this environment, the RL agent receives a $7 \times 7 \times 3 = 147$-dimensional vector as observation input and must select an action from the action space $\mathcal{A}$, which comprises six options. Notably, the $147$-dimensional vector serves as a compact and efficient representation of the environment, as opposed to directly representing the real pixels. In this work, we consider $8$-qubit systems, the $147$-dimensional vector is transformed into $8$-dimensional vectors using a simple classical linear layer, represented by the $H_{\delta}$ function in \sectionautorefname{\ref{sec:DiffQAS_QRL}}. These $8$-dimensional vectors are then processed by the VQC. The classical linear layer and the entire DiffQAS units (depicted in \figureautorefname{\ref{fig:DiffStructure_VQC}}) are trained together in an end-to-end manner.
The action space $\mathcal{A}$ comprises six actions {$0$,$\cdots$,$5$} available for the agent to select. These actions include \textit{turn left}, \textit{turn right}, \textit{move forward}, \textit{pick up an object}, \textit{drop the object being carried}, and \textit{toggle}. However, it is noteworthy that only the first three actions have tangible effects in the scenarios explored in this study. The agent is tasked with learning this distinction.
Within this environment, the agent garners a reward of 1 upon successfully attaining the goal. However, a penalty is deducted from this reward following the formula $1 - 0.9 \times (\textit{number of steps}/\textit{max steps allowed})$, where the maximum permissible step count is delineated as $4 \times n \times n$, with $n$ representing the grid size \cite{gym_minigrid}. This reward mechanism poses a challenge due to its \emph{sparse} nature, wherein rewards are only dispensed upon goal achievement.
As depicted in \figureautorefname{\ref{fig:minigrid_envs}}, the agent, denoted by the red triangle, is tasked with determining the most direct route from the initial position to the designated goal, depicted in green.
We consider six cases in this environment: MiniGrid-Empty-5x5-v0 (\figureautorefname{\ref{fig:minigrid_envs}}a), MiniGrid-Empty-6x6-v0 (\figureautorefname{\ref{fig:minigrid_envs}}b), MiniGrid-Empty-8x8-v0 (\figureautorefname{\ref{fig:minigrid_envs}}c), MiniGrid-SimpleCrossingS9N1-v0 (\figureautorefname{\ref{fig:minigrid_envs}}d), MiniGrid-SimpleCrossingS9N2-v0 (\figureautorefname{\ref{fig:minigrid_envs}}e) and MiniGrid-SimpleCrossingS9N3-v0 (\figureautorefname{\ref{fig:minigrid_envs}}f). Here the $N$ represents the number of valid crossings across walls from the starting position to the goal.
\begin{table}[htbp]
\caption{VQC baselines.}
\label{tab:different_circuit_baselines}
\begin{tabular}{|l|l|l|l|l|l|l|}
\hline
\diaghead{\theadfont Diag ColumnmnHead II}%
  {Component}{VQC config}                        & 1 & 2 & 3 & 4 & 5 & 6 \\ \hline
Encoding                & $R_{y}$  & $R_{z}$  & $R_{z}$  & $R_{y}$  & $R_{x}$  & $R_{x}$  \\ \hline
Trainable Rotation Gate & $R_{y}$  & $R_{y}$  & $R_{z}$  & $R_{z}$  & $R_{z}$  & $R_{y}$  \\ \hline
\end{tabular}
\end{table}
\subsection{Results}
\begin{figure}[htbp]
\begin{center}
\includegraphics[width=1\columnwidth]{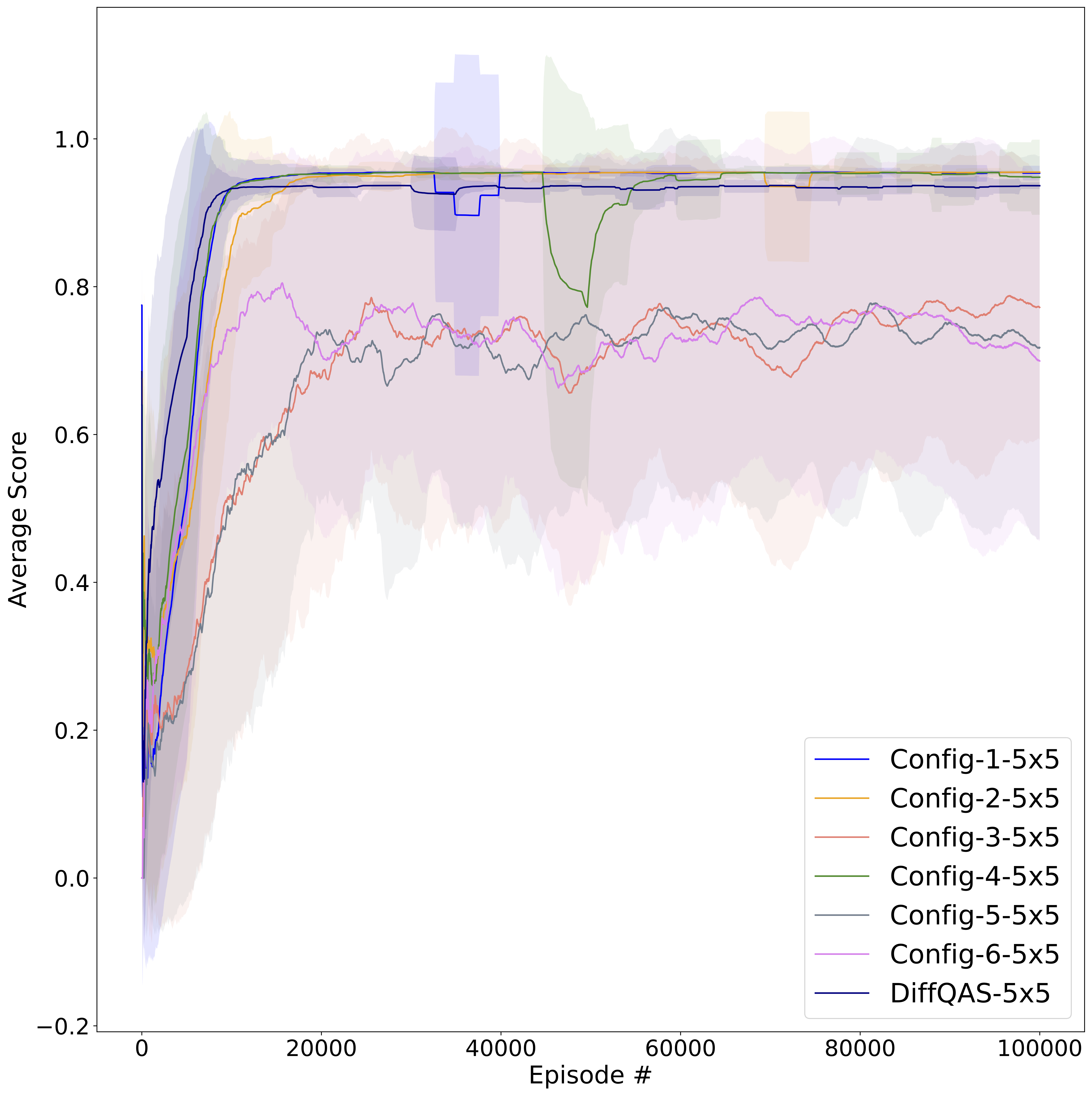}
\caption{{\bfseries Results: MiniGrid-Empty-5x5.}}
\label{fig:res_MiniGrid_Empty_5x5}
\end{center}
\end{figure}
In the environment \texttt{MiniGrid-Empty-5x5} (shown in \figureautorefname{\ref{fig:res_MiniGrid_Empty_5x5}}), we can see that the proposed DiffQAS can reach performance similar to the manually designed models (Config-1, Config-2 and Config-4). We can also observe that our DiffQAS model training is more stable regarding the average scores. In addition, our method can outperform the results from other manually designed models such as Config-3, Config-5 and Config-6 with a significant margin.
\begin{figure}[htbp]
\begin{center}
\includegraphics[width=1\columnwidth]{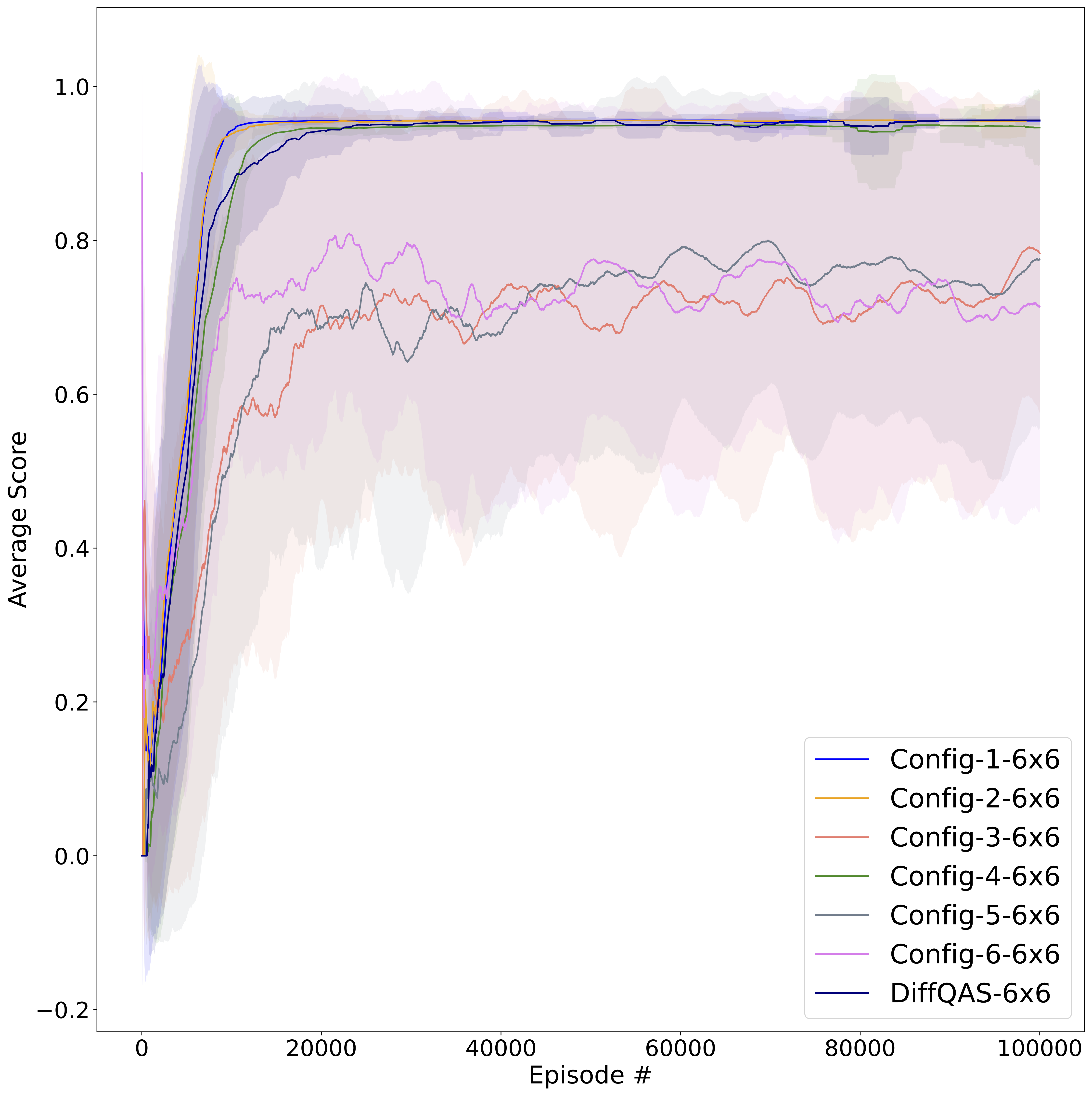}
\caption{{\bfseries Results: MiniGrid-Empty-6x6.}}
\label{fig:res_MiniGrid_Empty_6x6}
\end{center}
\end{figure}
In the environment \texttt{MiniGrid-Empty-6x6} (shown in \figureautorefname{\ref{fig:res_MiniGrid_Empty_6x6}}), we can see that the proposed DiffQAS can reach performance similar to the manually designed models (Config-1, Config-2 and Config-4). The proposed DiffQAS model converges a little bit slower than the manually-crafted Config-1 and Config-2, however, the DiffQAS model has no issue to reach the optimal score. The slower convergence may due to the larger search for the structural weights such that the program requires some additional time to learn these parameters. In addition, our method can outperform the results from other manually designed models such as Config-3, Config-5 and Config-6 with a significant margin.
\begin{figure}[htbp]
\begin{center}
\includegraphics[width=1\columnwidth]{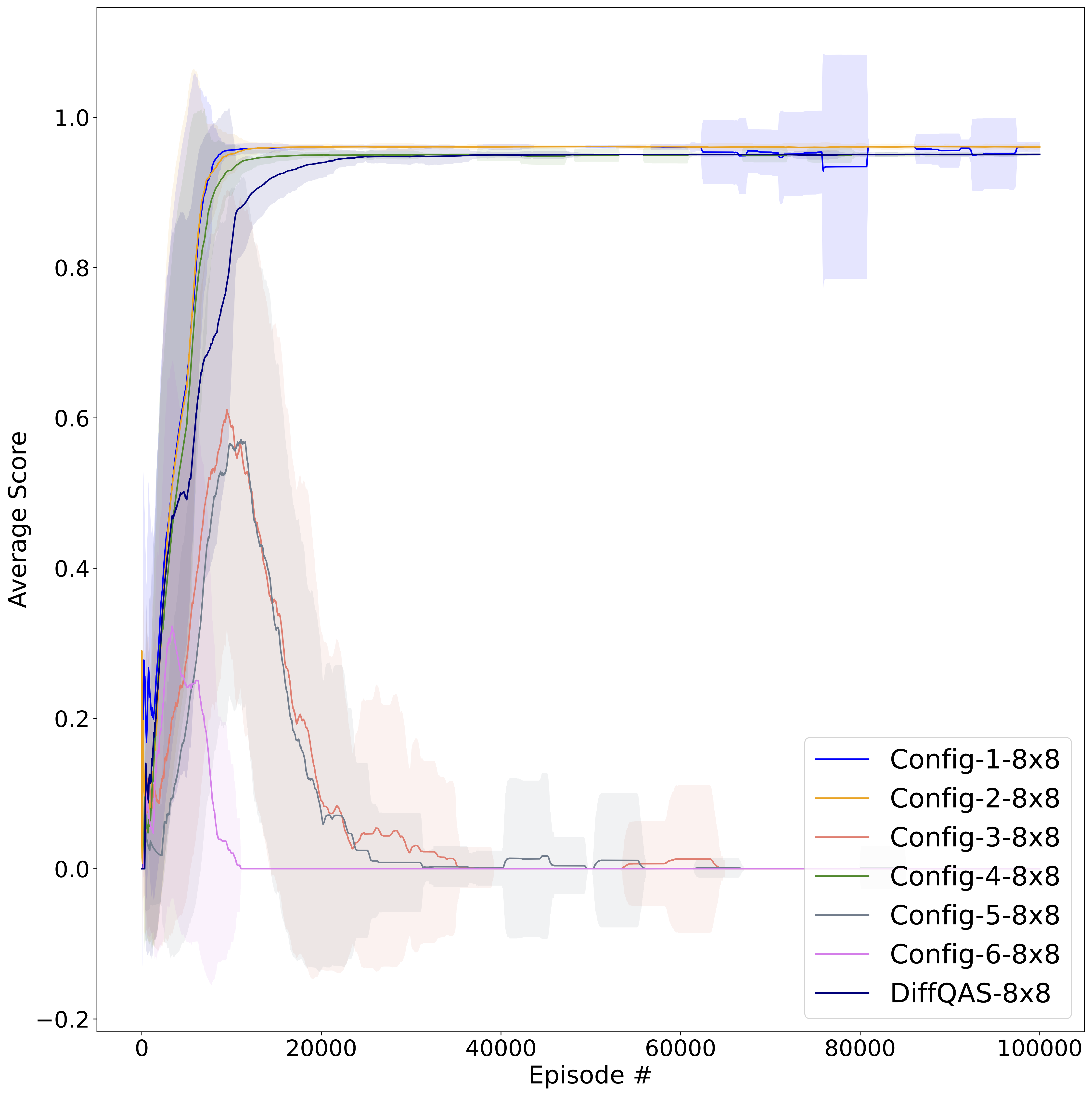}
\caption{{\bfseries Results: MiniGrid-Empty-8x8.}}
\label{fig:res_MiniGrid_Empty_8x8}
\end{center}
\end{figure}
In the environment \texttt{MiniGrid-Empty-8x8} (shown in \figureautorefname{\ref{fig:res_MiniGrid_Empty_8x8}}), we can see that the proposed DiffQAS can reach performance similar to the manually designed models (Config-1, Config-2 and Config-4). The proposed DiffQAS model converges a little bit slower than the three manually-crafted, however, the DiffQAS model has no issue to reach the optimal score and maintain the stability. The slower convergence may due to the larger search for the structural weights such that the program requires some additional time to learn these parameters. In addition, we can observe that the other three manually-crafted models Config-3, Config-5 and Config-6 fail to learn the policy at all. The environment \texttt{MiniGrid-Empty-8x8} is considered to be more difficult than the previous two, thus it it not surprising that certain models which perform poorly in previous case fail to learn the policy in this case.
\begin{figure}[htbp]
\begin{center}
\includegraphics[width=1\columnwidth]{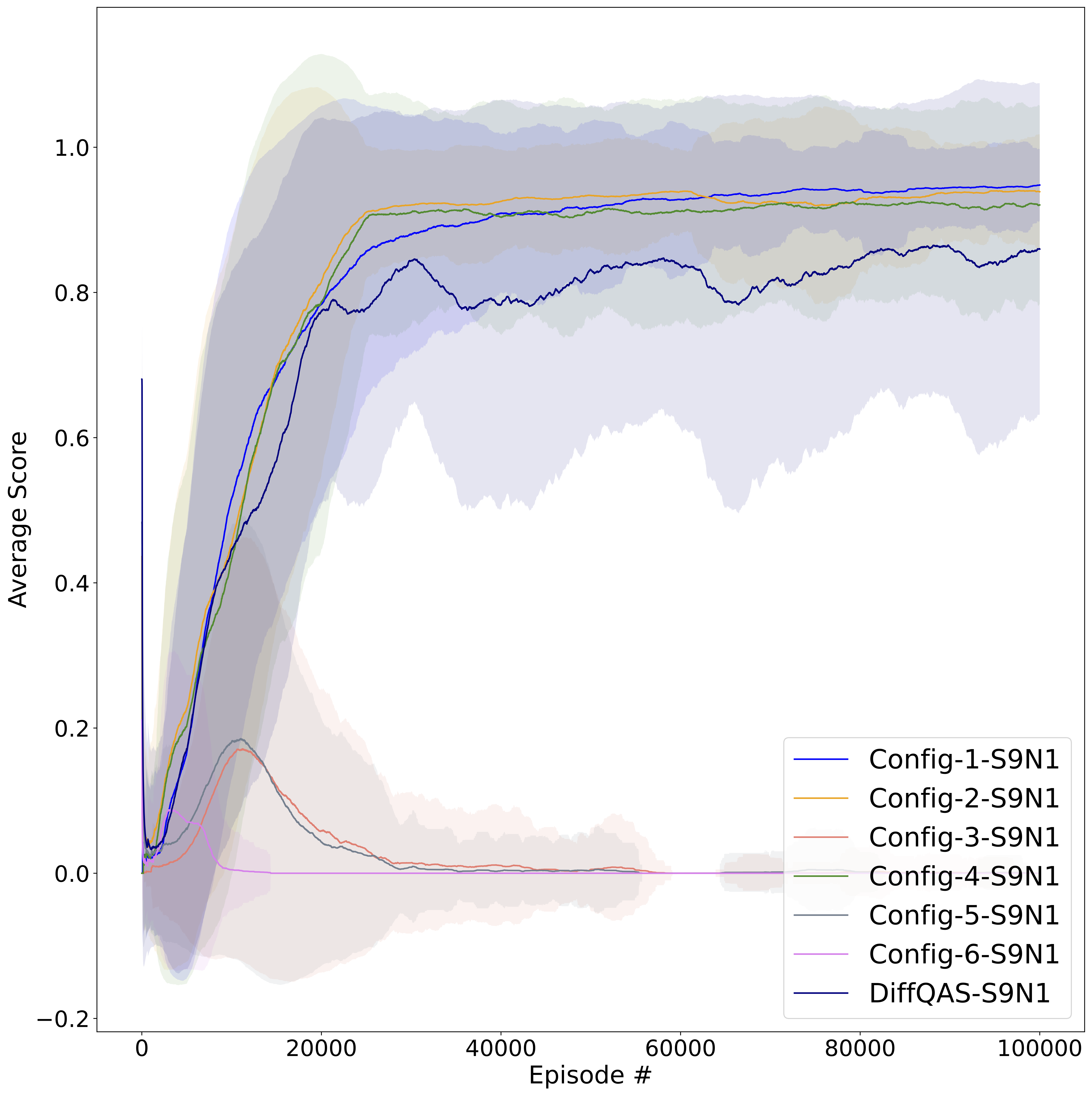}
\caption{{\bfseries Results: MiniGrid-SimpleCrossing-S9N1.}}
\label{fig:res_MiniGrid_SimpleCrossing_S9N1}
\end{center}
\end{figure}
In the environment \texttt{MiniGrid-SimpleCrossing-S9N1} (shown in \figureautorefname{\ref{fig:res_MiniGrid_SimpleCrossing_S9N1}}), we can see that the proposed DiffQAS can reach performance close to the manually designed models (Config-1, Config-2 and Config-4). The proposed DiffQAS model converges a bit slower than the three manually-crafted before the end of 100,000 training episodes. The slower convergence may due to the larger search for the structural weights such that the program requires some additional time to learn these parameters. In addition, we can observe that the other three manually-crafted models Config-3, Config-5 and Config-6 fail to learn the policy at all. The environment \texttt{MiniGrid-SimpleCrossing-S9N1} is considered to be more difficult than the previous \texttt{MiniGrid-Empty} environments, thus it it not surprising that certain models which perform poorly in previous cases fail to learn the policy in this case. 
\begin{figure}[htbp]
\begin{center}
\includegraphics[width=1\columnwidth]{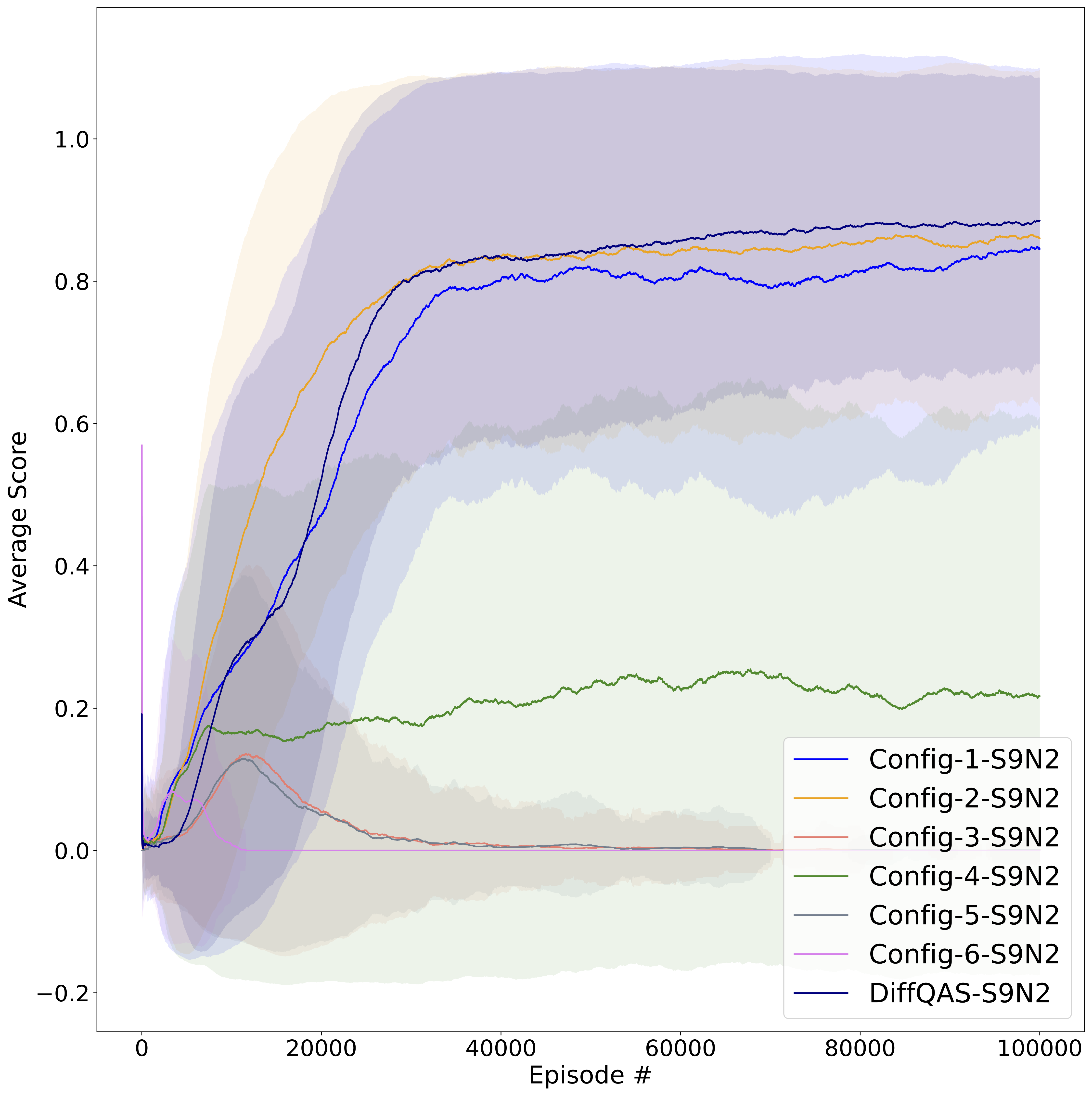}
\caption{{\bfseries Results: MiniGrid-SimpleCrossing-S9N2.}}
\label{fig:res_MiniGrid_SimpleCrossing_S9N2}
\end{center}
\end{figure}
In the environment \texttt{MiniGrid-SimpleCrossing-S9N2} (shown in \figureautorefname{\ref{fig:res_MiniGrid_SimpleCrossing_S9N2}}), we can see that the proposed DiffQAS can reach performance similar to the best manually designed models (Config-2) and beat all other manually-designed models. The proposed DiffQAS learns a bit slower than the best manually-crafted in early training episodes but finally surpass it with higher scores. The slower convergence may due to the larger search for the structural weights such that the program requires some additional time to learn these parameters before reaching optimal architecture weights. One of the previously good-preforming model (Config-4) now fails to learn the optimal policy in this more difficult environment. In addition, we can observe that the other three manually-crafted models Config-3, Config-5 and Config-6 fail to learn the policy at all. The environment \texttt{MiniGrid-SimpleCrossing-S9N2} is considered to be more difficult than the previous \texttt{MiniGrid-Empty} and \texttt{MiniGrid-SimpleCrossing-S9N1} environments, thus it it not surprising that certain models which perform poorly in previous cases fail to learn the policy in this case.
\begin{figure}[htbp]
\begin{center}
\includegraphics[width=1\columnwidth]{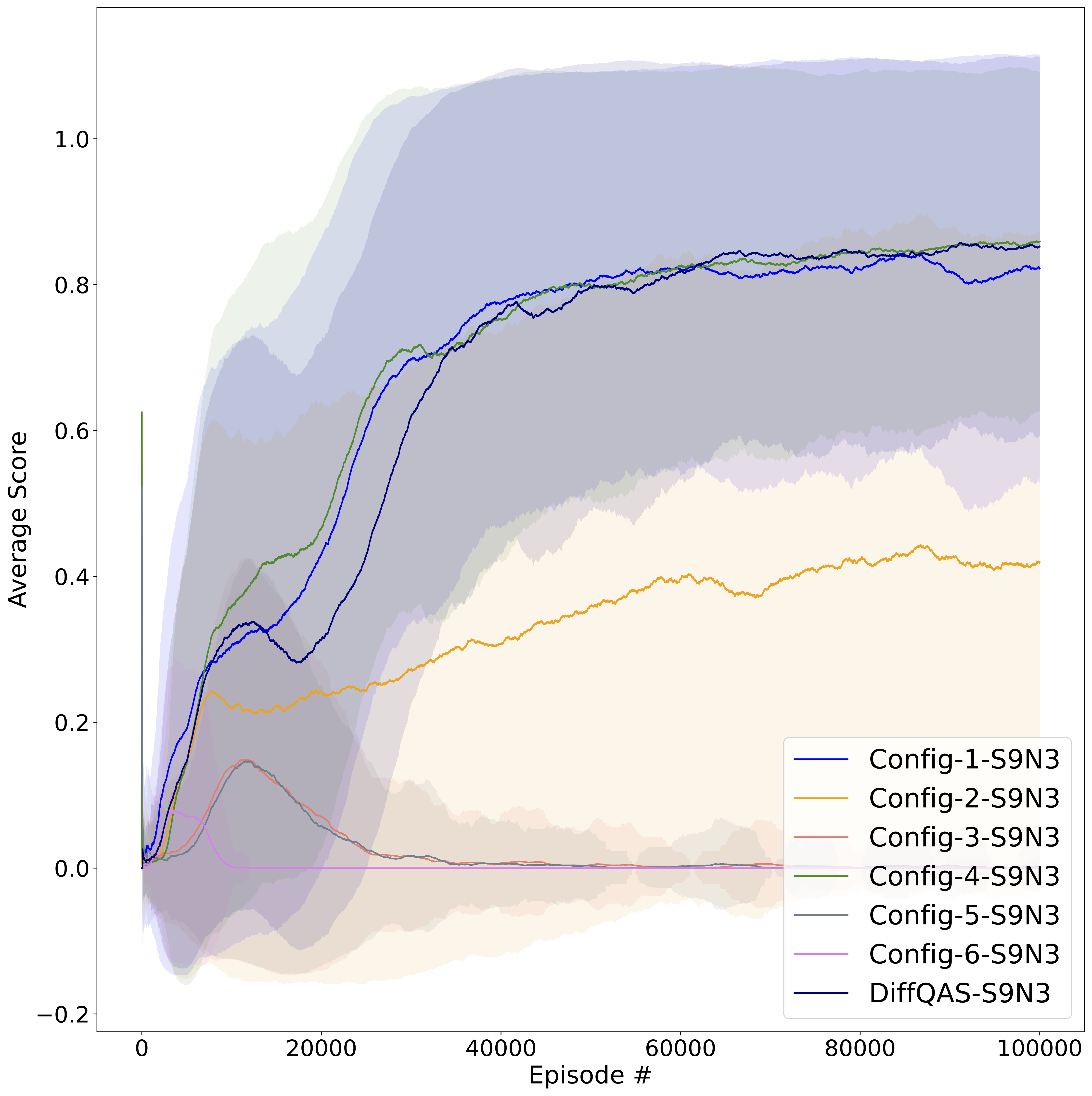}
\caption{{\bfseries Results: MiniGrid-SimpleCrossing-S9N3.}}
\label{fig:res_MiniGrid_SimpleCrossing_S9N3}
\end{center}
\end{figure}
In the environment \texttt{MiniGrid-SimpleCrossing-S9N3} (shown in \figureautorefname{\ref{fig:res_MiniGrid_SimpleCrossing_S9N3}}), we can see that the proposed DiffQAS can reach performance similar to the best manually designed models (Config-4) and beat all other manually-designed models at the end of training. The proposed DiffQAS learns a bit slower than the best manually-crafted in early training episodes but finally reaches the optimal scores. The observed slower convergence during early training episodes may due to the larger search for the structural weights such that the program requires some additional time to learn these parameters before reaching optimal architecture weights. One of the previously good-preforming model in  (Config-2) now fails to learn the optimal policy in this more difficult environment. In addition, we can observe that the other three manually-crafted models Config-3, Config-5 and Config-6 fail to learn the policy at all. The environment \texttt{MiniGrid-SimpleCrossing-S9N3} is considered to be more difficult than the previous \texttt{MiniGrid-Empty},  \texttt{MiniGrid-SimpleCrossing-S9N1} and \texttt{MiniGrid-SimpleCrossing-S9N2} environments, thus it it not surprising that certain models which perform poorly in previous cases fail to learn the policy in this case. Note that certain manually-designed architectures such as Config-2 and Config-4 cannot perform well consistently across different environments. This phenomenon further confirms the requirement to have a systemic and automatic way to build VQC architectures.

\section{\label{sec:Conclusion}Conclusion}
This paper introduces the DiffQAS-QRL framework, amalgamating differentiable quantum architecture search (DiffQAS) with quantum reinforcement learning (QRL). Specifically, we explore the quantum variant of asynchronous advantage actor-critic (QA3C) RL, capitalizing on parallel computing resources to augment training efficiency. Through numerical simulations conducted within diverse testing environments, our proposed DiffQAS-QRL method demonstrates its capability to identify well-performing VQC architectures across various scenarios, surpassing manually designed models in certain environments. These findings underscore the potential of DiffQAS-QRL in autonomously discovering high-performing VQC architectures for QRL tasks, thereby charting a novel course in the broader domain of automatic QML.
%

\bibliographystyle{ieeetr}
\bibliography{apssamp,bib/qas,bib/qrl,bib/tool,bib/qc,bib/qml_examples,bib/rl,bib/nas}

\providecommand{\noopsort}[1]{}\providecommand{\singleletter}[1]{#1}%
\begin{thebibliography}{10}

\bibitem{nielsen2010quantum}
M.~A. Nielsen and I.~L. Chuang, {\em Quantum computation and quantum information}.
\newblock Cambridge university press, 2010.

\bibitem{cerezo2021variational}
M.~Cerezo, A.~Arrasmith, R.~Babbush, S.~C. Benjamin, S.~Endo, K.~Fujii, J.~R. McClean, K.~Mitarai, X.~Yuan, L.~Cincio, {\em et~al.}, ``Variational quantum algorithms,'' {\em Nature Reviews Physics}, vol.~3, no.~9, pp.~625--644, 2021.

\bibitem{bharti2022noisy}
K.~Bharti, A.~Cervera-Lierta, T.~H. Kyaw, T.~Haug, S.~Alperin-Lea, A.~Anand, M.~Degroote, H.~Heimonen, J.~S. Kottmann, T.~Menke, {\em et~al.}, ``Noisy intermediate-scale quantum algorithms,'' {\em Reviews of Modern Physics}, vol.~94, no.~1, p.~015004, 2022.

\bibitem{mitarai2018quantum}
K.~Mitarai, M.~Negoro, M.~Kitagawa, and K.~Fujii, ``Quantum circuit learning,'' {\em Physical Review A}, vol.~98, no.~3, p.~032309, 2018.

\bibitem{chen2021end}
S.~Y.-C. Chen, C.-M. Huang, C.-W. Hsing, and Y.-J. Kao, ``An end-to-end trainable hybrid classical-quantum classifier,'' {\em Machine Learning: Science and Technology}, vol.~2, no.~4, p.~045021, 2021.

\bibitem{chen2022quantumCNN}
S.~Y.-C. Chen, T.-C. Wei, C.~Zhang, H.~Yu, and S.~Yoo, ``Quantum convolutional neural networks for high energy physics data analysis,'' {\em Physical Review Research}, vol.~4, no.~1, p.~013231, 2022.

\bibitem{oh2020tutorial}
S.~Oh, J.~Choi, and J.~Kim, ``A tutorial on quantum convolutional neural networks (qcnn),'' in {\em 2020 International Conference on Information and Communication Technology Convergence (ICTC)}, pp.~236--239, IEEE, 2020.

\bibitem{qi2023qtnvqc}
J.~Qi, C.-H.~H. Yang, and P.-Y. Chen, ``Qtn-vqc: An end-to-end learning framework for quantum neural networks,'' {\em Physica Scripta}, vol.~99, 12 2023.

\bibitem{wu2022poster}
J.~Wu and Q.~Li, ``Poster: Scalable quantum convolutional neural networks for edge computing,'' in {\em 2022 IEEE/ACM 7th Symposium on Edge Computing (SEC)}, pp.~307--309, IEEE, 2022.

\bibitem{chen2022quantumLSTM}
S.~Y.-C. Chen, S.~Yoo, and Y.-L.~L. Fang, ``Quantum long short-term memory,'' in {\em ICASSP 2022-2022 IEEE International Conference on Acoustics, Speech and Signal Processing (ICASSP)}, pp.~8622--8626, IEEE, 2022.

\bibitem{chen2022reservoir}
S.~Y.-C. Chen, D.~Fry, A.~Deshmukh, V.~Rastunkov, and C.~Stefanski, ``Reservoir computing via quantum recurrent neural networks,'' {\em arXiv preprint arXiv:2211.02612}, 2022.

\bibitem{bausch2020recurrent}
J.~Bausch, ``Recurrent quantum neural networks,'' {\em Advances in neural information processing systems}, vol.~33, pp.~1368--1379, 2020.

\bibitem{chu2023iqgan}
C.~Chu, G.~Skipper, M.~Swany, and F.~Chen, ``Iqgan: Robust quantum generative adversarial network for image synthesis on nisq devices,'' in {\em ICASSP 2023-2023 IEEE International Conference on Acoustics, Speech and Signal Processing (ICASSP)}, pp.~1--5, IEEE, 2023.

\bibitem{stein2021qugan}
S.~A. Stein, B.~Baheri, D.~Chen, Y.~Mao, Q.~Guan, A.~Li, B.~Fang, and S.~Xu, ``Qugan: A quantum state fidelity based generative adversarial network,'' in {\em 2021 IEEE International Conference on Quantum Computing and Engineering (QCE)}, pp.~71--81, IEEE, 2021.

\bibitem{yang2021decentralizing}
C.-H.~H. Yang, J.~Qi, S.~Y.-C. Chen, P.-Y. Chen, S.~M. Siniscalchi, X.~Ma, and C.-H. Lee, ``Decentralizing feature extraction with quantum convolutional neural network for automatic speech recognition,'' in {\em ICASSP 2021-2021 IEEE International Conference on Acoustics, Speech and Signal Processing (ICASSP)}, pp.~6523--6527, IEEE, 2021.

\bibitem{li2023pqlm}
S.~S. Li, X.~Zhang, S.~Zhou, H.~Shu, R.~Liang, H.~Liu, and L.~P. Garcia, ``Pqlm-multilingual decentralized portable quantum language model,'' in {\em ICASSP 2023-2023 IEEE International Conference on Acoustics, Speech and Signal Processing (ICASSP)}, pp.~1--5, IEEE, 2023.

\bibitem{yang2022bert}
C.-H.~H. Yang, J.~Qi, S.~Y.-C. Chen, Y.~Tsao, and P.-Y. Chen, ``When bert meets quantum temporal convolution learning for text classification in heterogeneous computing,'' in {\em ICASSP 2022-2022 IEEE International Conference on Acoustics, Speech and Signal Processing (ICASSP)}, pp.~8602--8606, IEEE, 2022.

\bibitem{di2022dawn}
R.~Di~Sipio, J.-H. Huang, S.~Y.-C. Chen, S.~Mangini, and M.~Worring, ``The dawn of quantum natural language processing,'' in {\em ICASSP 2022-2022 IEEE International Conference on Acoustics, Speech and Signal Processing (ICASSP)}, pp.~8612--8616, IEEE, 2022.

\bibitem{stein2023applying}
J.~Stein, I.~Christ, N.~Kraus, M.~B. Mansky, R.~M{\"u}ller, and C.~Linnhoff-Popien, ``Applying qnlp to sentiment analysis in finance,'' in {\em 2023 IEEE International Conference on Quantum Computing and Engineering (QCE)}, vol.~2, pp.~20--25, IEEE, 2023.

\bibitem{chen2023quantum_LSTM_RL}
S.~Y.-C. Chen, ``Quantum deep recurrent reinforcement learning,'' in {\em ICASSP 2023-2023 IEEE International Conference on Acoustics, Speech and Signal Processing (ICASSP)}, pp.~1--5, IEEE, 2023.

\bibitem{chen2023efficientQRL_QRC}
S.~Y.-C. Chen, ``Efficient quantum recurrent reinforcement learning via quantum reservoir computing,'' in {\em ICASSP 2024-2024 IEEE International Conference on Acoustics, Speech and Signal Processing (ICASSP)}, pp.~13186--13190, IEEE, 2024.

\bibitem{chen2022variationalQRL}
S.~Y.-C. Chen, C.-M. Huang, C.-W. Hsing, H.-S. Goan, and Y.-J. Kao, ``Variational quantum reinforcement learning via evolutionary optimization,'' {\em Machine Learning: Science and Technology}, vol.~3, no.~1, p.~015025, 2022.

\bibitem{chen19}
S.~Y.-C. Chen, C.-H.~H. Yang, J.~Qi, P.-Y. Chen, X.~Ma, and H.-S. Goan, ``Variational quantum circuits for deep reinforcement learning,'' {\em IEEE Access}, vol.~8, pp.~141007--141024, 2020.

\bibitem{lockwood2020reinforcement}
O.~Lockwood and M.~Si, ``Reinforcement learning with quantum variational circuit,'' in {\em Proceedings of the AAAI Conference on Artificial Intelligence and Interactive Digital Entertainment}, vol.~16, pp.~245--251, 2020.

\bibitem{skolik2021quantum}
A.~Skolik, S.~Jerbi, and V.~Dunjko, ``Quantum agents in the gym: a variational quantum algorithm for deep q-learning,'' {\em Quantum}, vol.~6, p.~720, 2022.

\bibitem{jerbi2021variational}
S.~Jerbi, C.~Gyurik, S.~Marshall, H.~Briegel, and V.~Dunjko, ``Parametrized quantum policies for reinforcement learning,'' {\em Advances in Neural Information Processing Systems}, vol.~34, pp.~28362--28375, 2021.

\bibitem{dong2008quantum}
D.~Dong, C.~Chen, H.~Li, and T.-J. Tarn, ``Quantum reinforcement learning,'' {\em IEEE Transactions on Systems, Man, and Cybernetics, Part B (Cybernetics)}, vol.~38, no.~5, pp.~1207--1220, 2008.

\bibitem{hsiao2022unentangled}
J.-Y. Hsiao, Y.~Du, W.-Y. Chiang, M.-H. Hsieh, and H.-S. Goan, ``Unentangled quantum reinforcement learning agents in the openai gym,'' {\em arXiv preprint arXiv:2203.14348}, 2022.

\bibitem{lan2021variational}
Q.~Lan, ``Variational quantum soft actor-critic,'' {\em arXiv preprint arXiv:2112.11921}, 2021.

\bibitem{kolle2024quantum}
M.~K{\"o}lle, M.~Hgog, F.~Ritz, P.~Altmann, M.~Zorn, J.~Stein, and C.~Linnhoff-Popien, ``Quantum advantage actor-critic for reinforcement learning,'' {\em arXiv preprint arXiv:2401.07043}, 2024.

\bibitem{CHEN2023321Async}
S.~Y.-C. Chen, ``Asynchronous training of quantum reinforcement learning,'' {\em Procedia Computer Science}, vol.~222, pp.~321--330, 2023.
\newblock International Neural Network Society Workshop on Deep Learning Innovations and Applications (INNS DLIA 2023).

\bibitem{chen2024learning}
S.~Y.-C. Chen, ``Learning to program variational quantum circuits with fast weights,'' {\em arXiv preprint arXiv:2402.17760}, 2024.

\bibitem{kuo2021quantum}
E.-J. Kuo, Y.-L.~L. Fang, and S.~Y.-C. Chen, ``Quantum architecture search via deep reinforcement learning,'' {\em arXiv preprint arXiv:2104.07715}, 2021.

\bibitem{ye2021quantum}
E.~Ye and S.~Y.-C. Chen, ``Quantum architecture search via continual reinforcement learning,'' {\em arXiv preprint arXiv:2112.05779}, 2021.

\bibitem{kimura2022quantum}
T.~Kimura, K.~Shiba, C.-C. Chen, M.~Sogabe, K.~Sakamoto, and T.~Sogabe, ``Quantum circuit architectures via quantum observable markov decision process planning,'' {\em Journal of Physics Communications}, vol.~6, no.~7, p.~075006, 2022.

\bibitem{sogabe2022model}
T.~Sogabe, T.~Kimura, C.-C. Chen, K.~Shiba, N.~Kasahara, M.~Sogabe, and K.~Sakamoto, ``Model-free deep recurrent q-network reinforcement learning for quantum circuit architectures design,'' {\em Quantum Reports}, vol.~4, no.~4, pp.~380--389, 2022.

\bibitem{lu2023qas}
X.~Lu, K.~Pan, G.~Yan, J.~Shan, W.~Wu, and J.~Yan, ``Qas-bench: rethinking quantum architecture search and a benchmark,'' in {\em International Conference on Machine Learning}, pp.~22880--22898, PMLR, 2023.

\bibitem{kundu2024enhancing}
A.~Kundu, P.~Bede{\l}ek, M.~Ostaszewski, O.~Danaci, Y.~J. Patel, V.~Dunjko, and J.~A. Miszczak, ``Enhancing variational quantum state diagonalization using reinforcement learning techniques,'' {\em New Journal of Physics}, vol.~26, no.~1, p.~013034, 2024.

\bibitem{sunkel2023ga4qco}
L.~S{\"u}nkel, D.~Martyniuk, D.~Mattern, J.~Jung, and A.~Paschke, ``Ga4qco: genetic algorithm for quantum circuit optimization,'' {\em arXiv preprint arXiv:2302.01303}, 2023.

\bibitem{zhu2023quantum}
X.~Zhu and X.~Hou, ``Quantum architecture search via truly proximal policy optimization,'' {\em Scientific Reports}, vol.~13, no.~1, p.~5157, 2023.

\bibitem{chen2023QRL_QAS}
S.~Y.-C. Chen, ``Quantum reinforcement learning for quantum architecture search,'' in {\em Proceedings of the 2023 International Workshop on Quantum Classical Cooperative}, pp.~17--20, 2023.

\bibitem{selig2023deepqprep}
P.~Selig, N.~Murphy, D.~Redmond, and S.~Caton, ``Deepqprep: Neural network augmented search for quantum state preparation,'' {\em IEEE Access}, 2023.

\bibitem{sun2024quantum}
Y.~Sun, Z.~Wu, Y.~Ma, and V.~Tresp, ``Quantum architecture search with unsupervised representation learning,'' {\em arXiv preprint arXiv:2401.11576}, 2024.

\bibitem{ostaszewski2021reinforcement}
M.~Ostaszewski, L.~M. Trenkwalder, W.~Masarczyk, E.~Scerri, and V.~Dunjko, ``Reinforcement learning for optimization of variational quantum circuit architectures,'' {\em Advances in Neural Information Processing Systems}, vol.~34, pp.~18182--18194, 2021.

\bibitem{wang2023automated}
P.~Wang, M.~Usman, U.~Parampalli, L.~C. Hollenberg, and C.~R. Myers, ``Automated quantum circuit design with nested monte carlo tree search,'' {\em IEEE Transactions on Quantum Engineering}, 2023.

\bibitem{he2023gnn}
Z.~He, X.~Zhang, C.~Chen, Z.~Huang, Y.~Zhou, and H.~Situ, ``A gnn-based predictor for quantum architecture search,'' {\em Quantum Information Processing}, vol.~22, no.~2, p.~128, 2023.

\bibitem{deng2023progressive}
M.~Deng, Z.~He, S.~Zheng, Y.~Zhou, F.~Zhang, and H.~Situ, ``A progressive predictor-based quantum architecture search with active learning,'' {\em The European Physical Journal Plus}, vol.~138, no.~10, p.~905, 2023.

\bibitem{yao2022monte}
J.~Yao, H.~Li, M.~Bukov, L.~Lin, and L.~Ying, ``Monte carlo tree search based hybrid optimization of variational quantum circuits,'' in {\em Mathematical and Scientific Machine Learning}, pp.~49--64, PMLR, 2022.

\bibitem{duong2022quantum}
T.~Duong, S.~T. Truong, M.~Pham, B.~Bach, and J.-K. Rhee, ``Quantum neural architecture search with quantum circuits metric and bayesian optimization,'' in {\em ICML 2022 2nd AI for Science Workshop}, 2022.

\bibitem{wu2023quantumdarts}
W.~Wu, G.~Yan, X.~Lu, K.~Pan, and J.~Yan, ``Quantumdarts: differentiable quantum architecture search for variational quantum algorithms,'' in {\em International Conference on Machine Learning}, pp.~37745--37764, PMLR, 2023.

\bibitem{zhang2022differentiable}
S.-X. Zhang, C.-Y. Hsieh, S.~Zhang, and H.~Yao, ``Differentiable quantum architecture search,'' {\em Quantum Science and Technology}, vol.~7, no.~4, p.~045023, 2022.

\bibitem{sun2024differentiable}
Y.~Sun, J.~Liu, Y.~Ma, and V.~Tresp, ``Differentiable quantum architecture search for job shop scheduling problem,'' {\em arXiv preprint arXiv:2401.01158}, 2024.

\bibitem{fosel2021quantum}
T.~F{\"o}sel, M.~Y. Niu, F.~Marquardt, and L.~Li, ``Quantum circuit optimization with deep reinforcement learning,'' {\em arXiv preprint arXiv:2103.07585}, 2021.

\bibitem{he2022quantum}
Z.~He, C.~Chen, L.~Li, S.~Zheng, and H.~Situ, ``Quantum architecture search with meta-learning,'' {\em Advanced Quantum Technologies}, vol.~5, no.~8, p.~2100134, 2022.

\bibitem{he2022search}
Z.~He, J.~Su, C.~Chen, M.~Pan, and H.~Situ, ``Search space pruning for quantum architecture search,'' {\em The European Physical Journal Plus}, vol.~137, no.~4, p.~491, 2022.

\bibitem{chen2022efficient}
Q.~Chen, Y.~Du, Q.~Zhao, Y.~Jiao, X.~Lu, and X.~Wu, ``Efficient and practical quantum compiler towards multi-qubit systems with deep reinforcement learning,'' {\em arXiv preprint arXiv:2204.06904}, 2022.

\bibitem{ding2022evolutionary}
L.~Ding and L.~Spector, ``Evolutionary quantum architecture search for parametrized quantum circuits,'' in {\em Proceedings of the Genetic and Evolutionary Computation Conference Companion}, pp.~2190--2195, 2022.

\bibitem{zhang2023evolutionary}
A.~Zhang and S.~Zhao, ``Evolutionary-based searching method for quantum circuit architecture,'' {\em Quantum Information Processing}, vol.~22, no.~7, p.~283, 2023.

\bibitem{ding2023multi}
L.~Ding and L.~Spector, ``Multi-objective evolutionary architecture search for parameterized quantum circuits,'' {\em Entropy}, vol.~25, no.~1, p.~93, 2023.

\bibitem{subasi2023toward}
O.~Subasi, ``Toward automated quantum variational machine learning,'' {\em arXiv preprint arXiv:2312.01567}, 2023.

\bibitem{sun2023differentiable}
Y.~Sun, Y.~Ma, and V.~Tresp, ``Differentiable quantum architecture search for quantum reinforcement learning,'' in {\em 2023 IEEE International Conference on Quantum Computing and Engineering (QCE)}, vol.~2, pp.~15--19, IEEE, 2023.

\bibitem{zhang2021neural}
S.-X. Zhang, C.-Y. Hsieh, S.~Zhang, and H.~Yao, ``Neural predictor based quantum architecture search,'' {\em Machine Learning: Science and Technology}, vol.~2, no.~4, p.~045027, 2021.

\bibitem{du2022quantum}
Y.~Du, T.~Huang, S.~You, M.-H. Hsieh, and D.~Tao, ``Quantum circuit architecture search for variational quantum algorithms,'' {\em npj Quantum Information}, vol.~8, no.~1, p.~62, 2022.

\bibitem{sutton2018reinforcement}
R.~S. Sutton and A.~G. Barto, {\em Reinforcement learning: An introduction}.
\newblock MIT press, 2018.

\bibitem{williams1992simple}
R.~J. Williams, ``Simple statistical gradient-following algorithms for connectionist reinforcement learning,'' {\em Machine learning}, vol.~8, no.~3-4, pp.~229--256, 1992.

\bibitem{mnih2016asynchronous}
V.~Mnih, A.~P. Badia, M.~Mirza, A.~Graves, T.~Lillicrap, T.~Harley, D.~Silver, and K.~Kavukcuoglu, ``Asynchronous methods for deep reinforcement learning,'' in {\em International conference on machine learning}, pp.~1928--1937, PMLR, 2016.

\bibitem{liu2018darts}
H.~Liu, K.~Simonyan, and Y.~Yang, ``Darts: Differentiable architecture search,'' {\em arXiv preprint arXiv:1806.09055}, 2018.

\bibitem{bergholm2018pennylane}
V.~Bergholm, J.~Izaac, M.~Schuld, C.~Gogolin, C.~Blank, K.~McKiernan, and N.~Killoran, ``Pennylane: Automatic differentiation of hybrid quantum-classical computations,'' {\em arXiv preprint arXiv:1811.04968}, 2018.

\bibitem{gym_minigrid}
M.~Chevalier-Boisvert, L.~Willems, and S.~Pal, ``Minimalistic gridworld environment for openai gym.'' \url{https://github.com/maximecb/gym-minigrid}, 2018.

\end{thebibliography}

\end{document}